\documentclass{article}

\usepackage{graphicx}

\begin{document}

\title{Collisional cooling of internal rotation in MgH$^+$ ions  trapped with He atoms: Quantum modeling meets  experiments in Coulomb crystals}

\date{\today}

\setcounter{footnote}{1}
\author{L. Gonz{\'a}lez-S{\'a}nchez\footnote{Departamento de Qu{\'i}mica F{\'i}sica, University of Salamanca, Plaza de los Ca{\'i}dos sn, 37008 Salamanca, Spain},\and 
R. Wester\footnote{Institut f\"ur Ionenphysik und Angewandte Physik, Universitaet  Innsbruck, Technikerstr. 25, A-6020, Innsbruck, Austria},\and
F.A. Gianturco\setcounter{footnote}{0}\thanks{corresponding author: francesco.gianturco@uibk.ac.at} \footnotemark[3]}


\maketitle

\section{Abstract}

Using  the  {\it ab initio} computed  Potential Energy Surface (PES) for the electronic interaction of the MgH$^+$($^1\Sigma$) ion with the He($^1$S)
atom, we calculate the relevant state-changing rotationally inelastic collision cross sections from a quantum treatment of the multichannel scattering
problem. We focus on the quantum dynamics at the translationally low energies for  the present partners discussed in  the earlier, cold ion trap experiments (see below) which we wish to model in detail. The corresponding state-changing rates computed between the lower rotational states of the molecular ion are employed  to describe  the time-evolution kinetics followed by recent experiments on Coulomb-crystalized  MgH$^+$($^1\Sigma$), where the ions are rotationally cooled by micromotion tuning after the uploading into the trap of  He as a buffer gas. The present computational modeling of the final ions' rotational temperatures in the experiments turns out to agree very well with their observations and points at a fast equilibration between rotational and thermal temperatures of the ions.

\section{Introduction}

The detailed control (for a general overview see \cite{1}) and the active manipulation of the internal, as well as the external degrees of freedom of gas-phase molecules has been 
pursued and investigated by now for the best part of the last twenty years \cite{2,3,4} and the information obtained has been of great value for furthering advances in several
experimental fields. Hence, from the progress in methodology \cite{5}, to quantum information processing \cite{6}, to the quantum control of molecular reactions
and transformations \cite{7} and to the collecting of accurate data for chemical reactions in well defined states \cite{8}, a great deal of studies and
of computational modeling have been developed. They have involved a large variety of  ensembles of cold molecules which could be further experimentally interrogated to follow their time evolutions from well-defined initial conditions and to extract specific information on their state-to-state collision rate constants \cite{9}.

From  these many fields of investigation the research on the behaviour of cold molecules, whether neutral or  ionic species, has developed fairly rapidly by employing a wide variety of techniques which will not be further discussed in this study as they have been already presented various times in the current literature \cite{10,11,12,13,14,15,16}.

The cooling techniques for molecular ions have also developed to the point that has become realistically possible to work with ensembles of molecular ions
that are sympathetically cooled into a Coulomb crystallization through an efficient Coulomb interaction with laser-cooled atomic ions \cite{15}.
The above techniques have shown that the translational-cooling schemes are indeed very versatile in bringing the molecular ions  down to  the low
temperatures of the millikelvin \cite{15}, although the further passage to also achieve simultaneously
that extremely low-lying internal states are being the most populated in the cold traps has to be  designed and adapted to the specific molecules under study \cite{15}.

In  very recent analysis \cite{17,18}, a novel setting has been experimentally investigated whereby the usual Helium buffer-gas technique for the cooling of internal molecular degrees of freedom has been employed for MgH$^+$($^1\Sigma$) ions. The ions were previously trapped in a cryogenically cooled, linear, radio-frequency quadrupole trap and further traslationally 
cooled, through a Coulomb type interaction, with simultaneously trapped, laser-cooled atomic Mg$^+$ ions \cite{17}.
 It was found there that the interaction with the additional He buffer gas is chiefly employed for the cooling of the molecular
ion's internal degrees of freedom, thereby needing much lower gas densities( e.g. around 10$^{10}$ cm$^{-3}$) for the uploaded buffer atoms, which can therefore be four to five orders of
magnitude lower that in a typical buffer-gas cooling setting \cite{18}. The vibrational degree of freedom of the MgH$^+$ partner is known to be already
frozen out at room temperature, with $>$ 99\% probability for the molecular ion of being in its vibrational ground state. Hence, at the cryogenic temperatures of the Coulomb
crystallization it can be entirely disregarded when modeling of the present dynamics, so that the full rotational-state distributions of the cold molecules could be directly measured in the Coulomb trap \cite{17}.

In the present work we shall analyze in detail this specific  collisional cooling process which is involving the differently populated rotational states of MgH$^+$ 
when the He atoms of the buffer gas are uploaded into the trap  after the formation of the Coulomb-crystallized ions.

The following Section \ref{sec2}  will provide specific information on the potential energy surface (PES) we have computed for the MgH$^+$/He system and 
will further outline the quantum dynamics of the rotationally inelastic collision processes. The next Section \ref{sec3} will analyse the relevant state-changing
cross sections and use them in the ensuing Section \ref{sec4} to generate the state-changing rates at the temperature of the traps. The master equations describing the system's time
evolution as a function of various trap parameters/conditions will be presented and discussed in Section \ref{sec5}. The final Section \ref{sec6}
will summarize our present conclusions.

\section{Interaction forces and quantum dynamics}
\label{sec2}

Within the usual Born-Oppenheimer (B.O.) approximation that separates nuclear and electronic motions, the electronic interaction between the MgH$^+$($^1\Sigma$) molecular ion at its equilibrium geometry of 1.67 \AA\    and the He($^1S$) neutral atom is described by a 2D grid of points providing the $V(R,\theta)$ single potential energy
surface (PES). In our earlier work on the same system \cite{19,20}, we computed the points by using the coupled-cluster single and double excitations with noniterative corrections for the triple excitation, CCSD(T), initial expansion coupled with a complete basis set (CBS) extrapolation limit and 
starting with the augmented coupled-cluster polarized valence multipole (auf-cc-pVnZ) (with n=3,4,5) basis set series as implemented in the software
package GAUSSIAN08 \cite{21}. The employed Jacobi coordinates were the distance $R$ of the He atom from the center-of-mass (c.o.m.) of MgH$^+$ and
the angle $\theta$ between $R$ and the bond, $r_{eq}$, of the partner molecular ion. The angular values were varied between 0º and 180º in intervals of 10º. The radial coordinate was ranged from 1.7 \AA\ to 16.0 \AA\, generating a total of 1200 radial points for the full set of angles mentioned before. in the current work we have taken advantage of the previous set of computed points of the 2D grid but we have added several new points to better describe the short-range repulsive interaction over a broad range of angles. Thus, an additional set of 100 points was added to the previous 2D grid.

The marked anisotropy of the present PES was already extensively discussed earlier \cite{19,20} so we will not repeat here the same analysis. Suffices it to
say that the most attractive well of the overall interaction is located along a linear structure with the He atom approaching the Mg$^+$ side 
of the molecular partner. The same PES becomes increasingly more repulsive as the He atom approaches the partner from the H-atom end of the molecular
cation. Thus, the multipolar representation of the anisotropic interaction can be obtained from writing:

\begin{equation}
	\label{expansion}
	V(R,\theta | r_{eq}) = \sum^{\lambda_{max}}_\lambda V_\lambda (R|r_{eq})P_\lambda(\cos\theta)
\end{equation}

where:

\begin{equation}
	\label{lambda}
	V_\lambda(R | r_{eq}) = \int^1_{-1} V(R,\theta|r_{eq})P_\lambda(\cos\theta)\, \mathrm{d}\cos\theta
\end{equation}

Hence, the range of action of each $V_\lambda$ coefficient gives us indication on the strength and range of the anisotropy present in the computed PES: each 
coefficient, in fact, will be directly involved in the dynamical coupling of rotational states of the cation during the collisional inelastic processes
within the coulomb trap, as we shall discuss below.
We have reached good numerical convergence of expansion \ref{expansion} by extending the sum up to $\lambda_{max}=30$. 

\begin{figure}
	\includegraphics{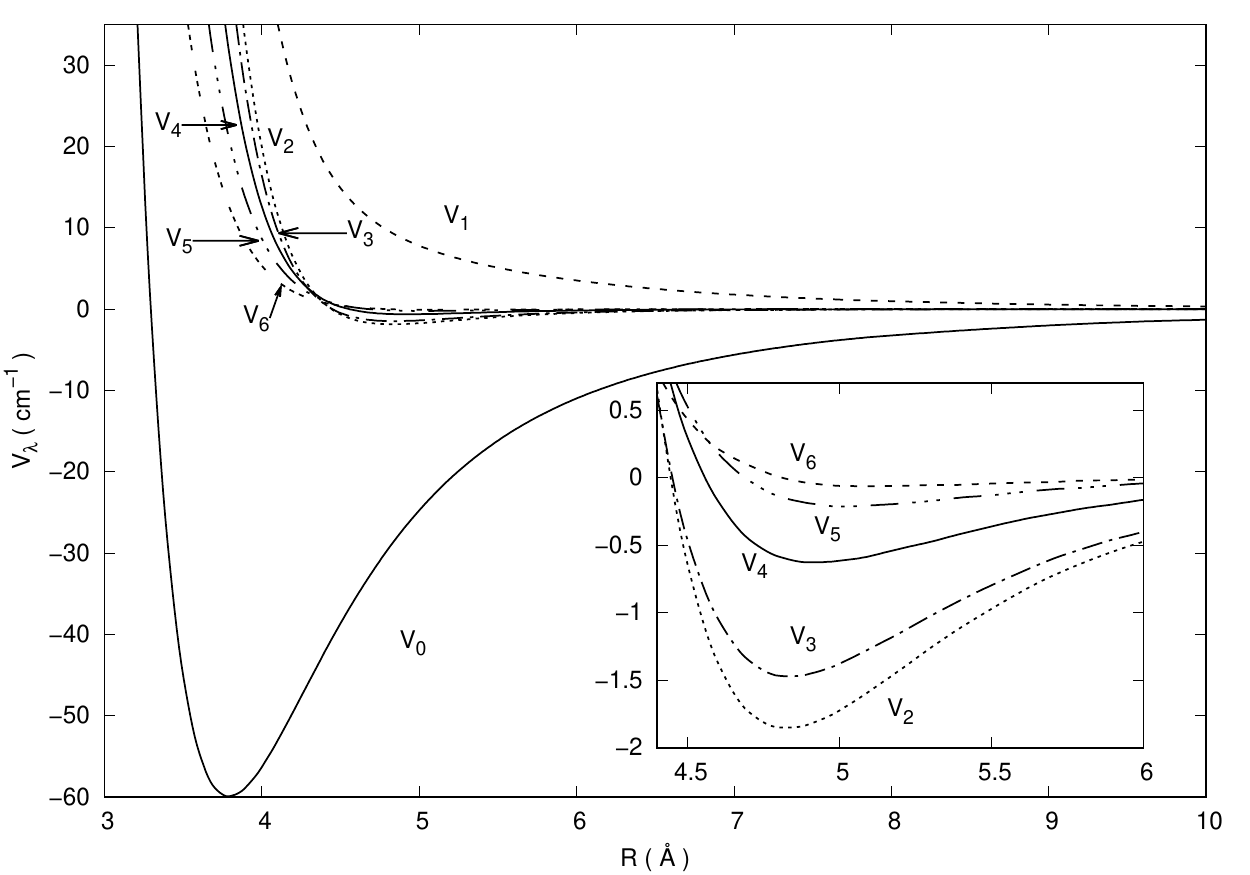}
	\caption{Computed multipolar coefficients from eq. \ref{lambda} for the MgH$^+$/He system. Only the first six coefficients are reported 
	in the figure.}
\label{fig1}
\end{figure}

As an example of the radial behaviour of the coefficients from eq. \ref{lambda}, we report in figure \ref{fig1} the first six coefficients for the
present PES.

The inset of this figure shows an enlarged view of the higher multipolar terms, beyond $\lambda$=0 and 1, in the radial regions where the coefficients
with $\lambda$ values = 2, 3, 4 and 5 present an attractive behaviour, albeit with decreasing depth as the $\lambda$ value increases,  while the next  higher term shows a much shallower well. All these terms, however,  are markedly less attractive than the dominant spherical term with $\lambda$=0  shown in the main figure.

Since  the data of figure \ref{fig1} mainly describe the short-range and the inner well regions, we need to further include the long-range (LR) behaviour of the total PES. This is done by a numerical interpolation between short-range (SR) and LR regions included in our in-house scattering code  (see below) and iturns out to show as its strongest attractive term   the long-range spherical polarizability tcontribution which appears in the standard treatment of the LR forces  via  perturbative expansions (e.g. see \cite{22}):

\begin{eqnarray}
	\label{vlambda3}
	V(R,\theta|r_{eq}) &\stackrel{R\rightarrow \infty}{=}& V_{LR}(R,\theta) \sim -\frac{\alpha_{He}}{2 R^4} - 2 \alpha_{He} 
	\frac{\mu P_1(\cos\theta)}{R^5} - \frac{\alpha_{He}\mu^2}{R^6} \\
	&&- \left(\alpha_{He}\mu^2 + Q\alpha_{He}\right)\frac{ P_2(\cos\theta)}{R^7}\\ + ...
	\label{perturbative}
\end{eqnarray}

The above array of asymptotic terms is dominated  by the spherical term  in the $\lambda$=0 coefficient. On the other hand, the coefficients with $\lambda$=1 and
$\lambda$=2 are next  in importance  with respect to the next  higher coefficients, as one could also gather from the relative strengths of their short-range terms shown 
by figure \ref{fig1}.
Hence, we could qualitatively say that rotational inelasticity at low collision energies will be mainly driven by the $\Delta j$=1 and 
$\Delta j$=2 rotational coupling terms of the PES, as we shall further discuss below.  

Once the full PES has been obtained and its multipolar coefficients generated from eq. \ref{expansion}-\ref{vlambda3}, including the coefficients
for the LR extension of the lowest three $\lambda$ values, one can then approach the calculations for the quantum multichannel dynamics of the
inelastic scattering processes inducing state-changes between rotational levels of the cation, taken to be in its $|v>$=0 vibrational
state: we therefore describe the nuclear motion of the partners within the usual time-independent Schr\"odinger equation (TISE) containing the 
potential interaction of eq. \ref{expansion} and subjected to the usual boundary conditions within the coupled-channel approach of expanding
the total wavefunctions on an ensemble of rotational functions for the molecular ion plus the  continuum functions for the relative
motion, numerically obtained at the positive, relative collision energies of the scattering partners \cite{23}.

We have employed our in-house numerical code ASPIN and details of its implementation have been given before \cite{24,25}. We will therefore not discuss
it again in the present work. Suffices it to say that the physical observables which we are obtaining from the ASPIN scattering code are in this case 
the state-to-state partial cross sections for each of the contributing total angular momenta $J$: $\sigma^J (j'\leftarrow j|E_i)$ with
$E_i$ giving the initial relative energy between partners. The further summing over the contributing angular momenta (which, in the present case
were taken up to $J_{max}$=50)  will therefore yield the corresponding state-to-state partial integral cross sections:
\begin{eqnarray}
	\label{xsec_def}
	\sigma(j'\leftarrow j|E_i) = \sum^{J_{max}}_J \sigma^J (j'\leftarrow j|E_i)
\end{eqnarray}
From them we can further obtain the partial rotational quenching/heating rate constants $K_{jj'}(T)$  at the temperature of interest:
\begin{eqnarray}
	\label{rate_def}
K_{jj'}(T) = \int \sigma(j'\leftarrow j|E) \sqrt{\frac{4 E}{\pi(k_B T)^3}} \exp{(-E/k_B T)} EdE
\end{eqnarray}
We have  integrated the computed cross sections over an  extended range of collision energy  for  the corresponding cross sections, ensuring that the threshold behaviour is well described by a dense grid of values. We have further  used and extended range of energies well beyond necessary extension to map the required interval of temperatures. Numerical convergence has been checked to a more than 0.01 stability of the final rates.

\section{Computing the state-changing collision cross sections}
\label{sec3}
As mentioned earlier, the inelastic cross sections were obtained using our in-house quantum CC code ASPIN,  \cite{25,26,27}. Therefore, we shall report here only a few specific details of the numerical procedure.  We have included in each CC calculation a maximum number of rotational channel up to $j_{max}$=11, where at each collision 
energy at least five channels were included as closed channels, to ensure overall convergence of the inelastic cross sections.

The radial integration was extended, at the lowest collision energies which we needed to take into consideration, out to $R_{max}$=1000 \AA. The anisotropy of the PES
was also included via a variable number of $\lambda$ values in the expansions of eqs. \ref{expansion} and \ref{lambda}. In practice, 
we found that we obtained converged inelastic cross sections by keeping $\lambda_{max}$=18 in our potential expansion. The $B_0$ value for the
MgH$^+$ rotor was taken to be 6.3870 cm$^{-1}$ \cite{30,31}. It is worth noting here that the present calculations cover a range of energies/temperatures which is much higher than that studied earlier by us on the present system \cite{19,20}. This therefore means that all the present cross sections were not among those discussed in that earlier work.

\begin{figure}
	\includegraphics{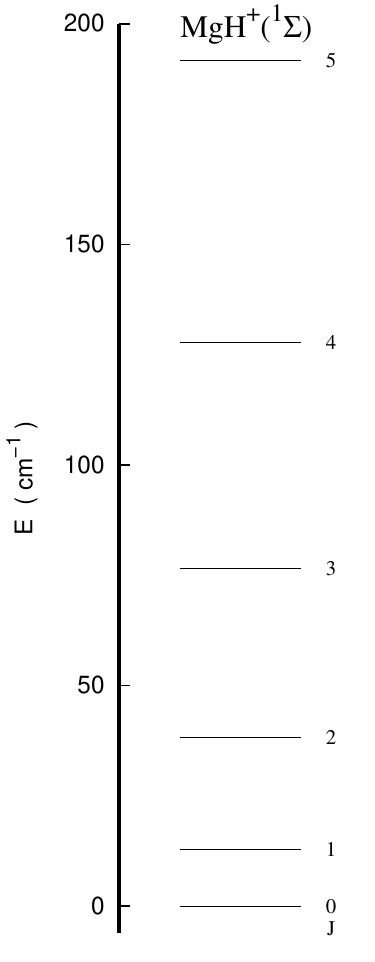}
	\caption{Computed rotational energy spacings between the lower five molecular levels of the ion which will be included
	in the present dynamical analysis of the collisional rotational energy transfer.}
\label{fig2}
\end{figure}

The data in figure \ref{fig2} show a pictorial presentation of the energy spacing for the lower rotational states of the MgH$^+$ cation
 considered in the present study.

One clearly sees in that figure how the lower three rotational levels are the closer in energy and will be the ones more effectively activated at 
the collisional temperatures of the present study. To reach  numerical convergence of the state-changing cross sections from eq. \ref{xsec_def}, 
however, we have included in the CC state expansion also the higher rotational levels shown by figure \ref{fig2}.

\begin{figure}
	\includegraphics{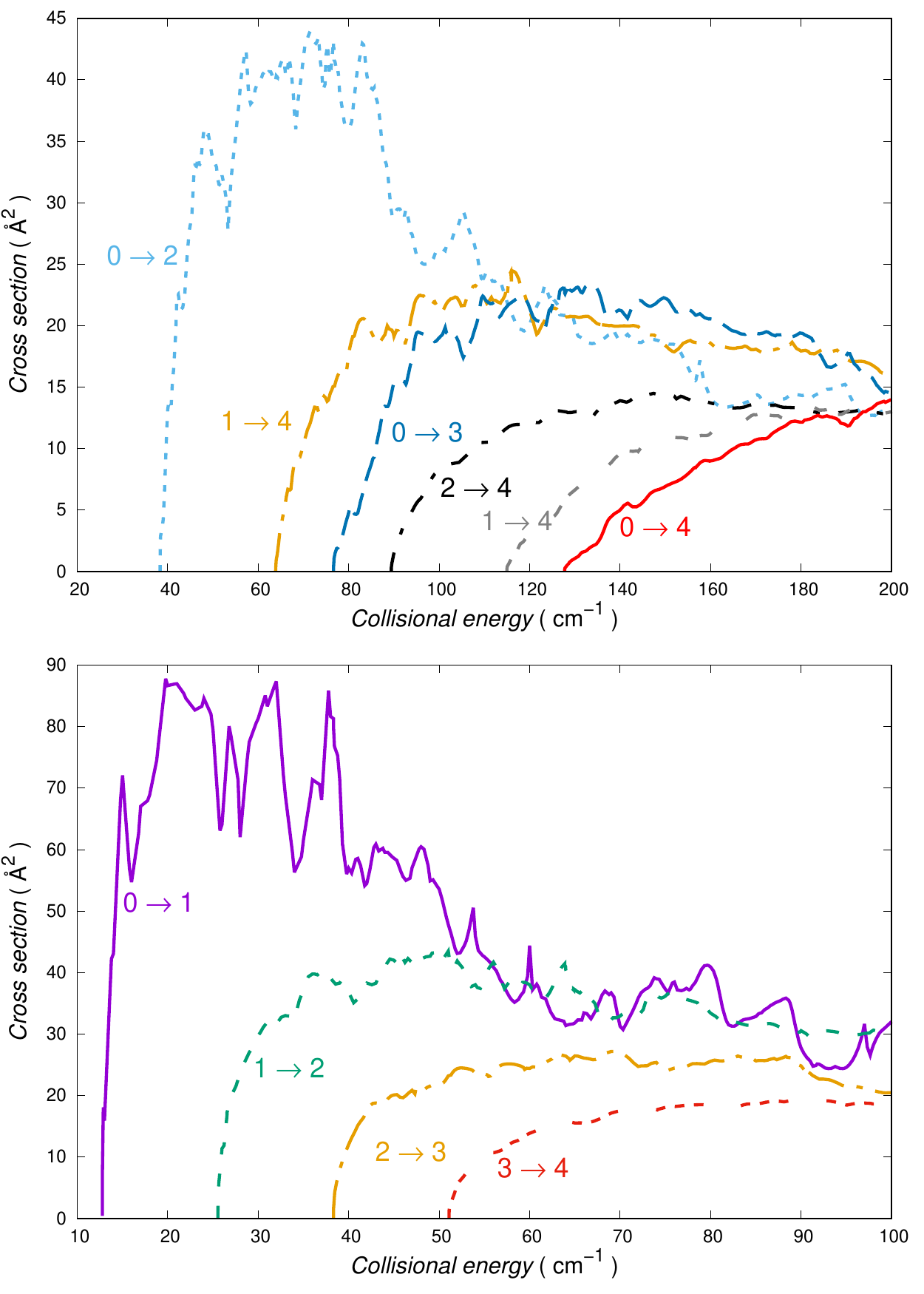}
	\caption{Computed excitation cross sections for collisional state-changing processes in the cold trap. The examined range of relative
	energies span 100 cm$^{-1}$. Upper panel: excitation processes for $\Delta j>$1 . Lower panel: excitation collisions
	for $\Delta j$=1 transitions.}
\label{fig3}
\end{figure}

The significant role played by anisotropic features of the PES, in conjunction with the values of the energy gaps between transitions, could be 
seen by the partial, excitation cross sections reported by the two panels of figure \ref{fig3}. The upper panel shows excitation processes with
$\Delta j>$1 transitions, while the processes involving $\Delta j$=1 transitions are shown by the lower panel.

The marked interaction forces which act during the collisional events are causing a very rich presence of resonant features, especially above
the onset energies of the excitation precesses and for transitions involving the lower rotational states.  Such marked structural features
are obviously linked to the occurrence of both open channel (trapping) resonances and to virtual excitations via Feshbach resonances involving 
closed rotational channels. The detailed analysis of the dynamical origins of such resonances will not be carried out here, as it is somewhat
outside the scope of the present study. However, it is interesting to point out that near each threshold the two dominant excitation processes
are those where the ground rotational state of the target ion is being excited via the $\Delta j$=1 coupling potential and the 
$\Delta j$=2 dynamical potential term . As discussed earlier from the PES features of figure \ref{fig1}, the largest cross sections pertain to the 
effects of the anisotropic coupling linked to the $\lambda$=1 multipolar coefficient,and to the extension of its radial range during the dynamics. On the other hand, 
 the next cross section in size is that for which it is the $\lambda$=2  potential coupling which chiefly causes the ($0\rightarrow 2$) rotational excitation process shown in the upper panel of the same figure \ref{fig3}. On the  whole, however, the data for the excitation processes show the state-changing dynamics to be an effective collisional path for the target ion at the low collision energies shown here.

\begin{figure}
	\includegraphics{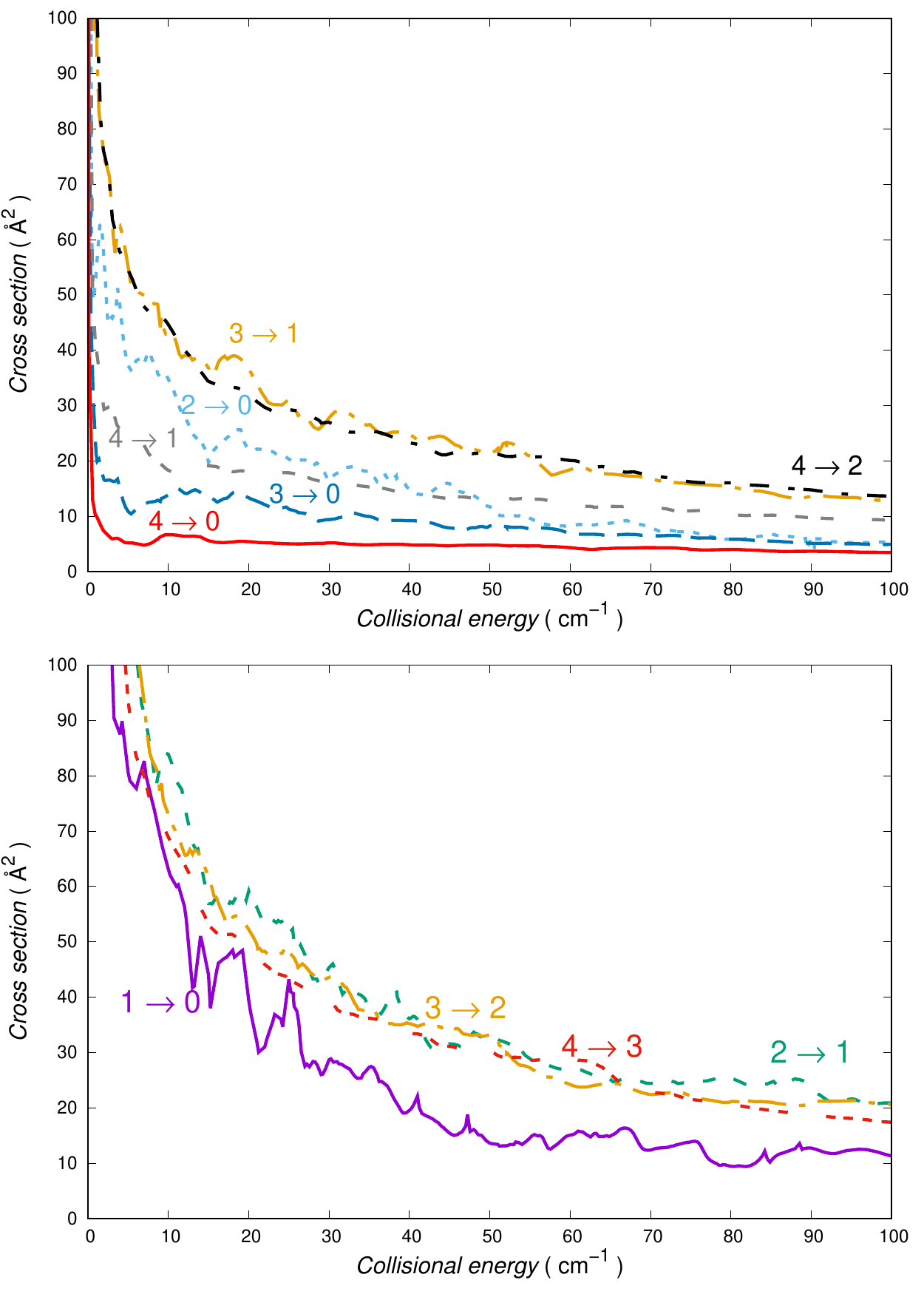}
	\caption{Computed rotationally inelastic cross sections associated with the de-excitation paths from the lower rotational states of the 
	molecular ion. Upper panel: rotational de-excitation processes with  $\Delta j>$1 ; lower panel: rotational "cooling" processes for $\Delta j$=1 transitions.
	See main text for further details.}
\label{fig4}
\end{figure}

The data reported by figure \ref{fig4} present the corresponding inelastic transitions relevant for the collisional rotational  "cooling" (i.e. rotational de-excitation) dynamics in the trap.
These de-excitation cross sections indicate that the relative sizes of the  inelastic transitions with $\Delta j$=-1
(lower panel in the figure) decrease as the initial state moves up on the energy ladder. It signifies that the processes where more 
internal rotational energy is released into the trap  after the collision are more efficient since the interaction times are shorter in comparison to the case where
the least energy is being released (e.g. for the ($1\rightarrow 0$) process) and the de-excitation dynamics is least effective. 
On the other hand, when the rotational internal energy released becomes even larger (upper panel of figure \ref{fig4}), the presence $\Delta j >$1 couplings
make the dynamical torques activated by the higher terms of the multipolar potential of eqs. \ref{expansion} less efficient so that all these inelastic cross sections 
are uniformly smaller than those with  $\Delta j$=1 state-changing processes. Here again, however, we see that the cross sections are rather significant in size and indicate
the inelastic collisions to be an effective path for depopulating the rotational states of the trapped molecular ions.

\section{Rotationally inelastic rates at low T ($\le$50K)}
\label{sec4}

Following the relation shown by equation \ref{rate_def} we have employed the inelastic cross sections discussed in the previous section
to obtain the corresponding inelastic rates over a broad range of temperatures, spanning those of the Coulomb-crystals experiments \cite{16}.

\begin{figure}
	\includegraphics{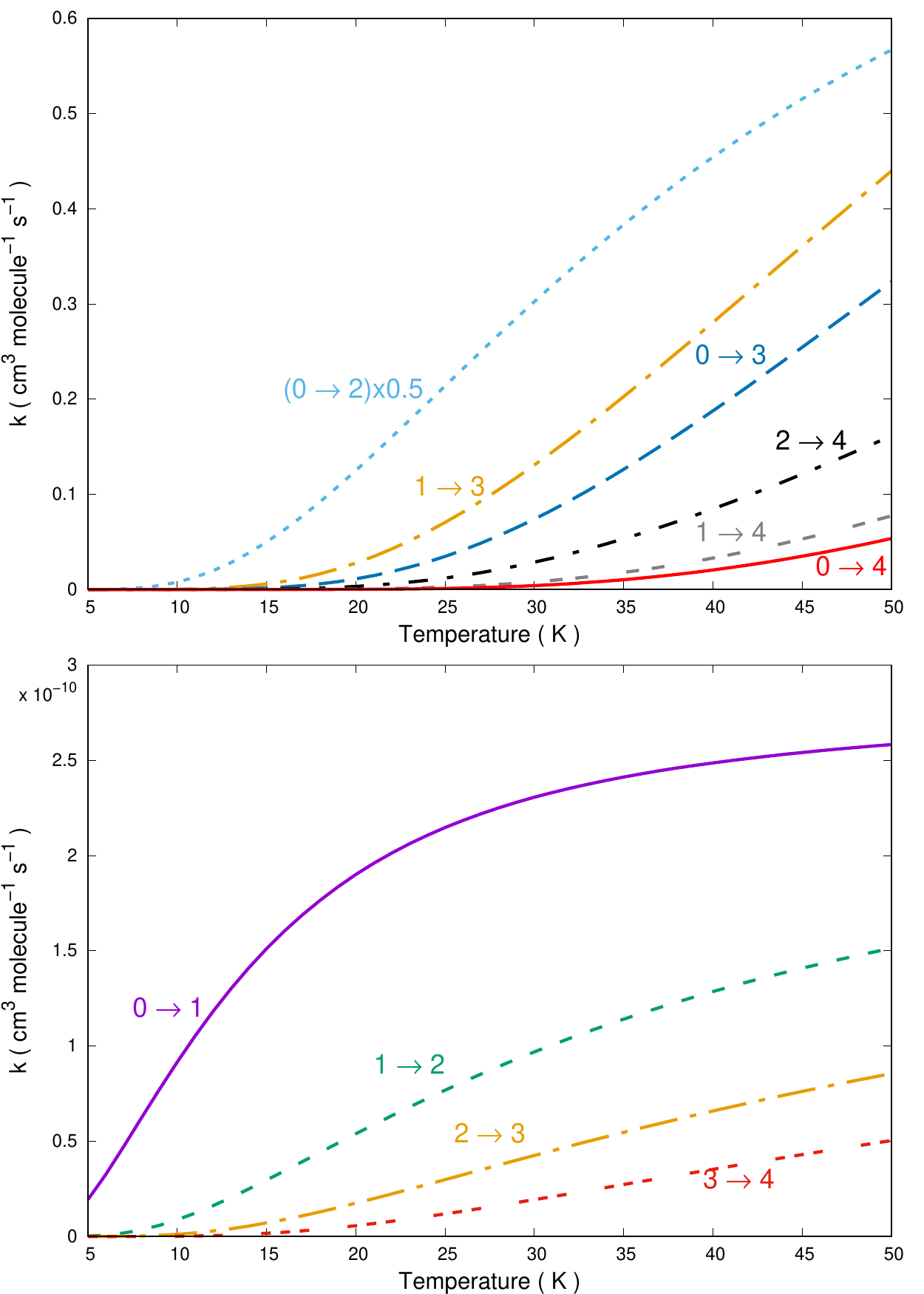}
	\caption{Computed rotational 'heating' (excitation) rates between different rotational states of MgH$^+$ in the traps. Upper panel: excitation rates with
	$\Delta j>$1. Lower panel: excitation rates for $\Delta j$=1 transitions. See main text for further details.}
\label{fig5}
\end{figure}

\begin{figure}
	\includegraphics{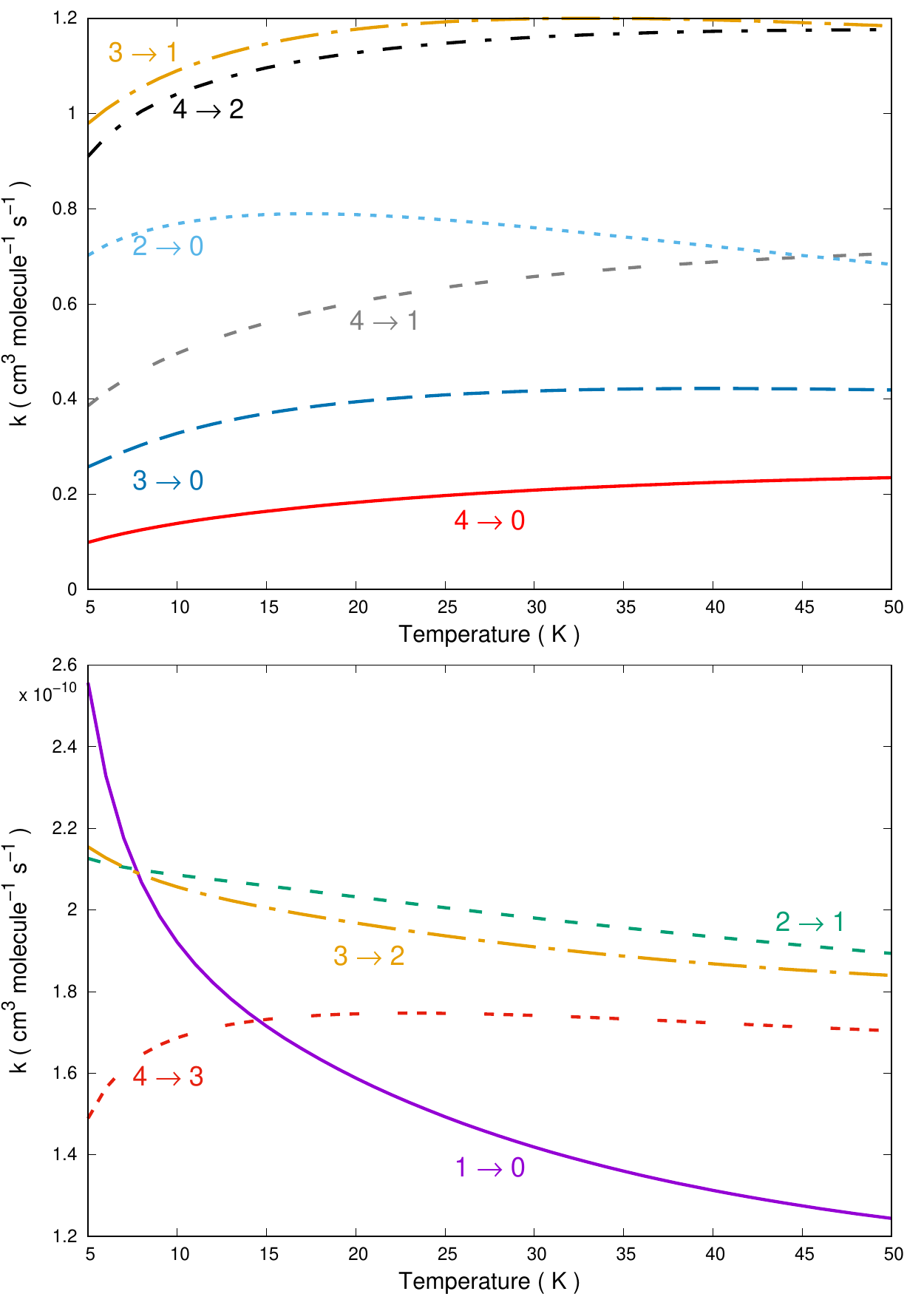}
	\caption{Same as in figure \ref{fig5}, but this time for the 'cooling'  (de-excitation) rates between rotational states. The upper panel reports 
	inelastic rates with a negative $\Delta j>$-1, while the lower panel shows transitions with negative $\Delta j$ values=-1. See main text for
	further details.}
\label{fig6}
\end{figure}

The data in figure \ref{fig5} and \ref{fig6} present the state-to-state inelastic rates involving the lower five rotational states of 
the MgH$^+$ trapped ion. More specifically, the excitation rates reported by figure \ref{fig5} show, in the upper panel of that figure, the processes for which the state-changing
transitions involve $\Delta j>$1 transitions, while the lower panel presents the transition rates with $\Delta j$=1 and involving rotational
states from $|0>$ up to $|4>$. The following consideration could be made by examining the results given in the figure:
\begin{enumerate}
	\item the cooling rates become rapidly fairly large as the collision energy increases and show the ($0\rightarrow 1$) excitation
		to be by far the largest. All other single-quantum excitations from the higher states of the ion show rates which are smaller than
		the ($0\rightarrow 1$) rate by factor of two or more;
	\item the multi-quantum excitation rates are seen to be fairly smaller than the former ones (see upper panel in the figure) and
		remain nearly one order of magnitude smaller in the temperature regions between 10K and 30K;
	\item in both sets of processes the excitation from the ground rotational state correspond to the largest values of the excitation
		rates within each panel.
\end{enumerate}

It is interesting to note that in a recent study of ours on a similar molecular cation, the OH$^+$($^2\Sigma$) molecule \cite{31},
the rotational excitation rates computed in a trap with He as a buffer gas, turned out to be about a factor of three smaller
over the same range of temperatures, in keeping with differences between the energy spacings of their lower-lying rotational levels.

If we now turn to the rotational relaxation rates over the same range of temperatures, we see in the two panels of figure \ref{fig6} their
relative size and T-dependence. The lower panel reports single-quantum rotational ''cooling'' transitions, while the multiple-quantum rotational de-excitation 
transitions are in  the upper panel of the same figures.

The general trend of the relative sizes of the inelastic rotational de-excitation transitions is here very similar to that of the excitation rates. However, with the
exception of the ($1\rightarrow 0$)  process, all the computed rates show a fairly slow dependence on  temperature, a result which
is in keeping with the same findings from our earlier calculations regarding the OH$^+$/He system \cite{31}. The single-quantum rotational relaxation rates are also 
uniformly larger than those for two- and three-quanta transitions, which are in some cases up to one order of magnitude smaller. The same size differences
were also found for the OH$^+$($^2\Sigma$) cation \cite{31}.

\section{Modeling the cooling kinetics in the trap}
\label{sec5}

As discussed in the Introduction , the experimental findings of ref. \cite{18} indicate that the He buffer-gas uploaded within the setting
of MgH$^+$ ions trapped in a cryogenically cooled linear, radio-frequency quadrupole trap, and already translationally cooled through Coulomb
interaction with atomic  Mg$^+$ ions, can cause the molecular ions to be cooled into their ground rotational state even through a low density
of He atoms of $\sim$ 10$^{10}$ cm$^{-3}$ is being present in the trap.In those experiments, in fact, the He temperature is essentially kept constant while the effective collision "temperature" is changed via the scaling of the average micromotion amplitude. In this way, the molecular ions experience different effective temperatures within the uploaded gas environment (see ref.  \cite{18} for further details).

Given the information we have obtained from the calculations presented in the previous Sections, we are now in a position to try and follow
the microscopic evolutions of the cation's rotational state populations by setting up the corresponding rate eq.s describing such 
evolution  as induced by collisional energy transfer with the uploaded He atoms in the trap \cite{28,29}:
\begin{equation}
	\label{eq6}
	\frac{d\mathbf{p}}{dt} = n_{He} \mathbf{k}(T)\cdot \mathbf{p}(t)
\end{equation}
where the quantity $n_{He}$ indicates the density of He atoms in the trap, the vector $\mathbf{p}(t)$ contains the time-evolving 
fractional rotational populations of the ion partner's rotational state, p$_ j(t)$, from the initial conditions at t=t$_{initial}$, and the
matrix  $\mathbf{k}(T)$  contains the individual k$_{i\rightarrow j}(T)$ rate coefficients at the temperature of the trap's conditions. 
Both the p(t$_{initial}$) values and the collisional temperature T of the trap corresponding to the mean collisional energy between the partners
are quantities to be specifically selected in each computational run and will be discussed in detail in the modelling examples presented below. In
the present study we shall disregard for the moment the inclusion of the state-changing rates due to spontaneous radiative processes in the trap.
These quantities are already known to be smaller than the collisionally-controlled rates between the lower rotational levels of such systems, as 
already shown by us in earlier studies \cite{29}, and are therefore not expected to have a significant effect under the present trap conditions
\cite{18}.

We have chosen the initial rotational temperature of the trap's ions to be at 400 K, so that the vector's components at t=t$_{initial}$ are given
by a Boltzmann distribution at that chosen temperature. This was done  in order to follow the kinetics evolution over an extended range of time 
and also test the physical reliability of our computed state-changing collisional rates. Obviously the actual kinetics evolution of physical interest 
in this study will be considered over the range of the much lower temperatures sampled by the experiments \cite{18}.

If the rate coefficients of the $\mathbf{K}(T)$ matrix satisfy the detailed balance between state-changing transitions, then as t$\rightarrow \infty$ 
the initial Boltzmann distribution will approach that of the effective equilibrium temperature of the uploaded buffer gas as felt by the ions in the Coulomb Crystals.
 These asymptotic solutions correspond to the steady-state conditions in the trap and can be obtained by solving the corresponding homogeneous form of eq. \ref{eq6} given
as: $d\mathbf{p} (t)/dt = 0$. We solved the homogeneous equations by using the singular-value  decomposition technique (SVD) \cite{28}, already 
employed by us in previous studies. 
The non-homogeneous equations \ref{eq6}, starting from our t$_{initial}$ 
of 400 K, were solved using the Runge-Kutta method for different  translational temperatures of the trap. 
Since the role of the He density is simply that of a scaling factor in the kinetics eq.s, 
we present in the figures only the actual value which was employed in the trap experiments \cite{18}.

\begin{figure}
	\includegraphics{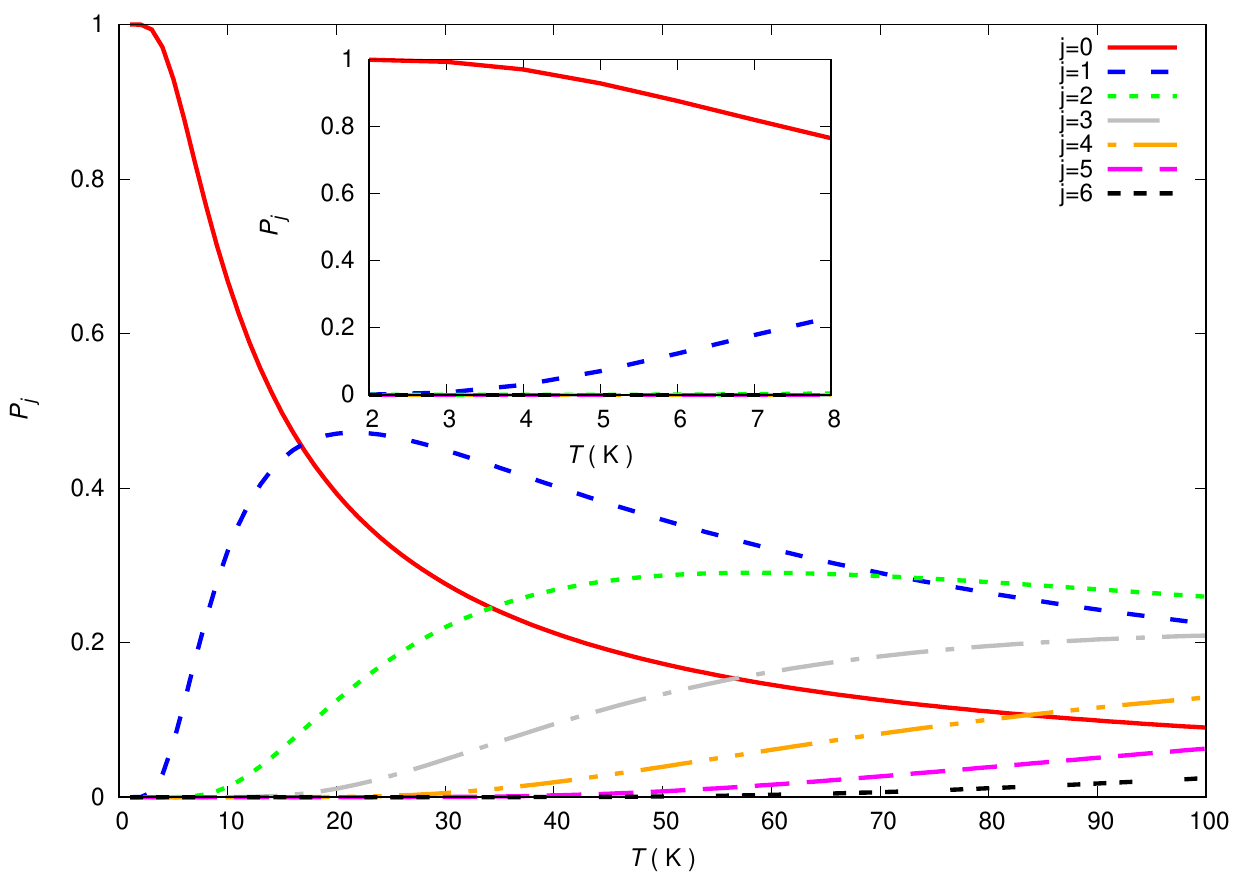}
	\caption{Asymptotic (steady-state) rotational populations of the MgH$^+$ internal states as a function of computed trap translational temperatures
	up to 100 K. The data for the lower T values up to $\sim$ 8 K is shown in the inset.  See main text for further details.}
\label{fig7}
\end{figure}

The results shown by figure \ref{fig7} indicate for the present system the steady-state population in the cold trap over a rather large
range of temperatures up to 100 K. It allows us to numerically control the equilibrium population expected for the rotational levels of the ions 
in the crystal as the perceived translational  temperature is increased. The inset in the figure clearly shows how, up to about 10 K, the only levels involved
would be j=0 and j=1, with a very small presence of the j=2 populations. The rather small energy spacings between the MgH$^+$ rotational
levels, for which the first four levels are spanning  about 75 cm$^{-1}$, indicates also that, as the temperature increases, many more 
rotational levels will be occupied at the equilibrium trap temperatures indicated by figure \ref{fig7}. How fast such a collisional thermalization
would occur will be shown by the results presented below. 

\begin{figure}
	\includegraphics{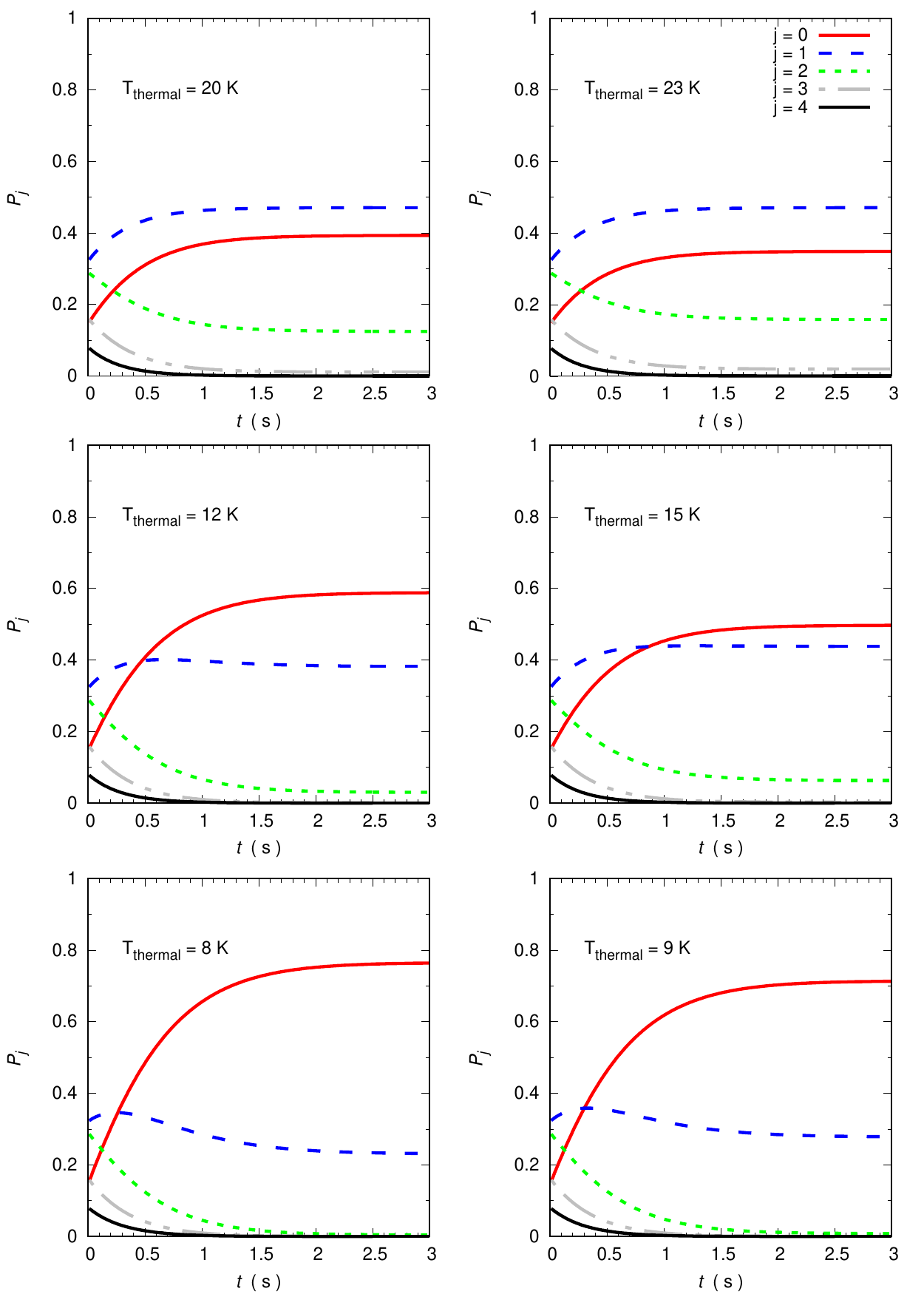}
	\caption{Computed values of the time-evolution  of the MgH$^+$ ion's rotational state populations by collisional perturbations in the Coulomb 
	crystals induced by tuning the micromotion amplitudes after uploading He as a buffer gas. The temperatures are the same as those sampled in 
the experiments of ref. \cite{18} and the gas density is also the same of the experiments: 10$^{10}$ cm$^{-3}$.}
\label{fig8}
\end{figure}

The calculations reported by figure \ref{fig8} indicate the time evolution of the collisionally-driven molecular ion populations for different 
values of the trap's temperature and for a fixed density of the buffer gas of $n_{He}$=10$^{10}$ cm$^{-3}$. All the temperatures reported in the panels
correspond  to those experimentally assessed in the Coulomb crystals of ref. \cite{18}, while the density of the He gas is also the one indicated by the 
experiments.

The six panels of that figure report the six different effective temperatures perceived by the localized ions  that are  reported by the experimental data \cite{18}.
We show  in the panels the time evolution for the relative populations of the lower five ion's rotational levels, although the experimental data analysed  the
behaviour of only  the first four rotational states as those having any significant population during the cooling process.

The data indicate that the change of the perceived trap temperature is indeed a significant parameter for changing the relative rates of level populations 
during the collisional rotational de-excitation processes. As T increases,  in fact, we see that the two dominant population fractions are those associated 
to the j=0 and j=1 states. Their relative importance, however, changes dramatically when moving from the 8-9 K region, where at the equilibrium time 
the j=0 population is about twice that of the j=1 state, to the 20-23 K region,
where the two levels have now inverted  populations and the j=1 state is more abundant than the j=0 state. This result is remarkably close to the experimental
findings \cite{18} which suggested that the rotational populations of the ion's levels gets very quickly to be that given by the thermal distributions
within the trap, at the average translational temperature generated by changes in the micromotion amplitude after the uploading of the buffer gas whose
 temperature remains fixed at around 8.7 K \cite {18}. We see, in fact, from the data of the calculations 
shown by the panes of figure \ref{fig8},
that after at most about 3 s the relative populations have reached their steady-state values reported by the test calculations of figure \ref{fig7}.
In other words, the efficient collisional energy transfer processes between MgH$^+$ and He, caused by changing their relative translational average energy in the trap
through the tuning of the micromotion amplitudes,  
are rapidly allowing the internal rotational populations
of the cation to thermalize at the experimentally achieved translational temperatures  in the Coulomb trap. The experimental data indeed 
suggest also (see figure 3 of ref. \cite{18}) that, over a fairly broad range of temperatures, the rotational temperatures are the same as the 
translational temperatures after  the He buffer gas is uploaded to the trap. It therefore stands to reason to expect, as found in the experiments, 
 that, after the passing of a very short time interval, the ions localized within the CC environment will have reached the same temperatures for their rotational and 
translational degrees of freedom, as we shall further illustrate below.

Another useful indicator which could be extracted from the present calculations is the definition of a characteristic time, $\tau$, which can 
be defined as:
\begin{equation}
	\label{tau}
	\left\langle E_{rot} \right\rangle (\tau) - \left\langle E_{rot} \right\rangle (t=\infty) = 
	\frac{1}{e}\left( \left\langle (E_{rot} \right\rangle (t=0) -\left\langle E_{rot} \right\rangle(t=\infty)\right)
\end{equation}
the quantity $\left\langle E_{rot} \right\rangle$ represents the level-averaged rotational internal energy of the molecule in the trap after a 
characteristic time interval $\tau$ defined by equation \ref{tau}. It obviously depends on the physical collision frequency and therefore
it depends on the $n_{He}$ value present in the trap. 

\begin{figure}
	\includegraphics{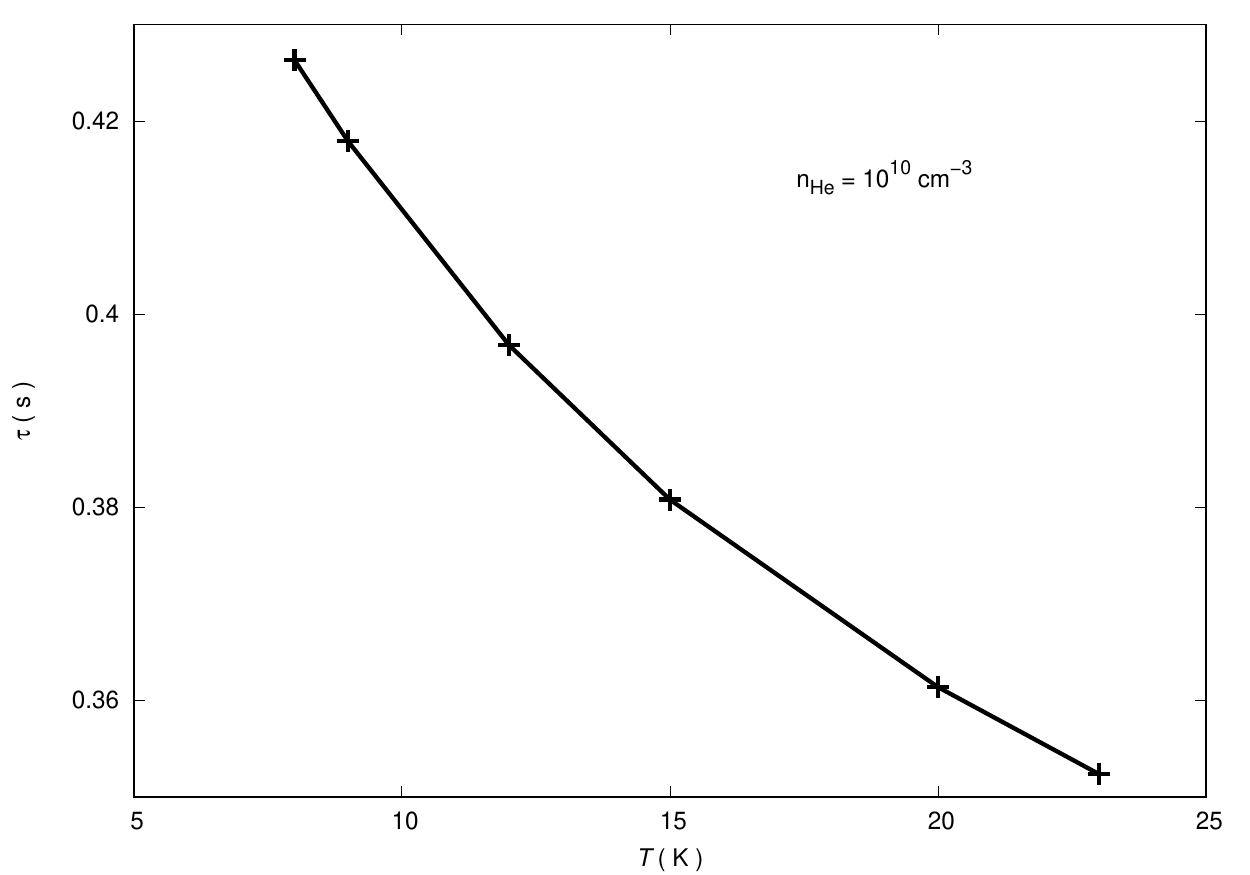}
	\caption{Computed characteristic relaxation time $\tau$, as defined in eq. \ref{tau}, for different translational (thermal) temperatures 
	and for the  buffer gas density value considered in the experiments.} 
\label{fig9}
\end{figure}

The model calculations of figure \ref{fig9} report the behaviour of $\tau$ for the experimental value of the He density in the trap  and for the expected range of 
effective thermal temperatures tested by the experiments \cite{18}.

From the data reported in that figure, we see that $\tau$ is a slow function of T, while it depends markedly on the chosen $n_{He}$ value and is
inversely proportional to it.  The buffer gas density is the one estimated by the experiments \cite{18} and the range of temperatures covers the values given
by the experimental data of figure 1 in \cite{18}. One clearly sees there that the characteristic relaxation time, i.e. the average elapsed time required to reach the rotational-to-translational temperatures,is well below 1 s, being around 0.50 s at the lower T values and only reducing to 0.30 s at the highest experimental thermal temperatures. 
Such values are once more indicative of the collisional efficiency of the rotational cooling processes for
MgH$^+$, since similar calculations for the OH$^+$($^2\Sigma$) cation \cite{31} indicated a $\tau$ value which was a factor of two larger over  the
same range of temperatures.

\begin{figure}
	\includegraphics{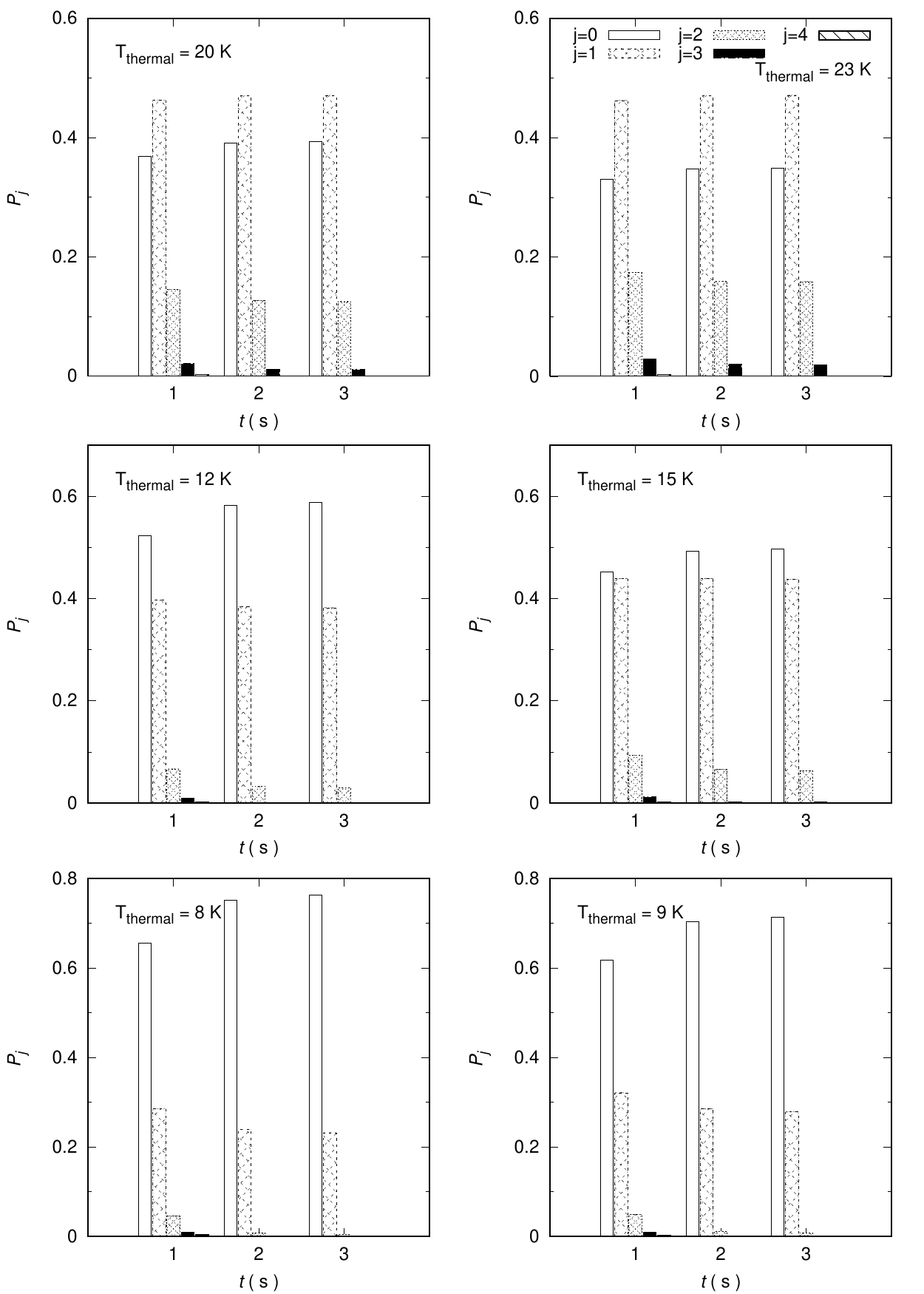}
	\caption{Computed relative populations of the MgH$^+$ rotational levels in the trap, as a function of trap's thermal temperatures and for three different values ( in sec) 
         of  time delay after buffer gas uploading and ion's micromotion tuning in the experiments.}
\label{fig10}
\end{figure}

To further make contact with the experimental findings, and link our present  results with the $\tau$ indicator of figure \ref{fig9}, we report in
figure \ref{fig10} the relative populations of the ion's rotational levels in the trap, as a function of the different temperatures sampled by the
experiments and for different delay times after the uploading of the buffer gas in the trap and the start of the micromotion scaling  to change the effective, relative  translational energies  within the trap. The time values shown correspond to 1 , 2 and 3 s of delay after buffer gas loading.

The following considerations could be made by looking at the results shown in that figure:
\begin{enumerate}
	\item All panels in the figure indicate that, after 3 s at the most, the population of rotational states has reached the Boltzmann's
		thermal distribution for each of the examined temperatures. This can be confirmed by the time evolutions in figure \ref{fig8} and
		the Boltzmann's distributions of figure \ref{fig7}. this means that the rotational temperature of the molecular ions is in therml equilibrium with its velocity distributions.
	\item With the exclusion of the two lowest temperatures of 8 K and 9 K, at all the other temperatures the relative populations change
		negligibly after increasing the time delays from 1 s to 3 sec. In practice, the calculations indicate that after about 1 s the
		rotational populations have reached their steady-state value at that temperature in each panel, while only at the lowest T values
		the equilibration  of the relative populations is reached after a slightly longer time (see also lowest two panels of figure \ref{fig8}).
\end{enumerate}

\begin{figure}
	\includegraphics[width=0.6\textwidth]{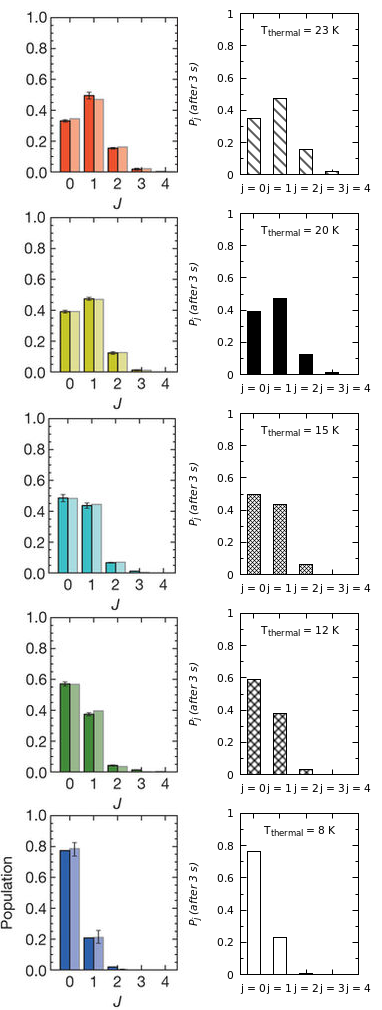}
	\caption{ Observed relative populations of the rotational states, and the observed thermal distributions of the  same,(given by the lighter 'sticks'), for
 the trapped MgH$^+$ cation in the Coulomb crystal  after the uploading of the He buffer gas (panels in the left column, reproduced with permission from ref. \cite{18}). 
We also report  for comparison the  calculated rotational population distributions after  solution of the master
	eq.s \ref{eq6} and after a time evolution delay of 3 s (panels on the right-side column of figure). See main text for further details.}
\label{fig11}
\end{figure}

One can therefore argue that the present collision-driven rotational population evolution in the Coulomb traps indicates a very rapid thermalization
process and a very efficient energy redistribution within the MgH$^+$ rotational levels  in order to bring the rotational temperature of 
the trapped ion in line with the translational temperature. The latter is the one  achieved by the same CC ions after the micromotion tuning of the relative collision energies following 
the uploading of the He as the  buffer gas. Thus, the ions change their relative velocities with respect to the He gas atoms but rapidly attain rotational stabilization in equilibrium with their final thermal energy.
To make the comparison with the experimental findings ever more transparent, we further report in the five panels of figure \ref{fig11}
the relative distributions of rotational state populations found by the five temperatures considered by the experiments (fig. 1 of ref. \cite{18})
and compare them with the same distributions found by us after a time delay of 3 s in the evolution of the kinetics eq.s. After what has been discussed above, 
one should also note  the rapid termalization of the molecular rotational temperatures to the ion's translational temperatures found by our calculations, a feature
which confirms the experimental findings reported by fig.1 in ref. \cite{18} and shown by the panels on the r.h.s. of figure \ref{fig11}.

One can make the following considerations from a perusal of the data presented in that figure:

\begin{enumerate}
	\item The experimental distributions at the various rotational temperatures we are considering here are very close to those obtained from solving the present
		Master eq.s and extracting from the latter  the distributions after an uploading time between 1 s and 3 s;
	\item The calculations suggest a very rapid, collision-driven thermalization between the internal rotational degrees of freedom
		of the trapped MgH$^+$ cation and the translational temperature experimentally generated in the trap after  the uploading of the buffer gas and the micromotion
                  tuning effects.  This result is in line with what has been suggested by the experimental data reported by \cite{18};
         \item  We also see from the experimental data reported in the figure that the best-fit thermal distributions given by the lighter "sticks" are very close
                      to the rotational temperatures reported in the same panels. This is in keeping with our present findings, since we have shown that the computed the rotational 
                      distributions on the r.h.s. of the figure where obtained after thermalization of the ion's rotational  temperatures to its steady-state translational temperatures;
	\item The experimental estimates of the refilling rates for the depleted rotational levels in the trap which are given in 
		ref. \cite{18} indicate a time of about 1 s$^{-1}$, from which they extrapolate a rate of about 10 s$^{-1}$ for a gas
		density $n_{He}$=10$^{10}$ cm$^{-3}$. This means that the driving depletion rates, over the range of T spanned
		by the experiments, would be around 10$\times$10$^{-10}$ cm$^3$s$^{-1}$. If we look at the individuals state-to-state,
		depletion rates computed in the present study (see panels of figure \ref{fig6}) we see that our dominant rates for 
		depleting the first three rotational states of MgH$^+$ around 10-20 K sum up to about 6.5$\times$10$^{-10}$ cm$^3$s$^{-1}$.
		This is in good accord with the above value, especially if we notice that the Langevin rate value employed by the experimental work
		is larger than the state-to-state cooling rates generated by the present calculations.Thus, we expect that the rate value extracted from the refilling frequencies
                  should be smaller when our computed rates would be employed. Either way, however, computed and experimentally estimated value are close to each other;
	\item One should further notice that our calculated estimates of the characteristic cooling time $\tau$ of figure \ref{fig9} are
		around 0.4 s at the temperature of 15 K. This corresponds to a ''refilling'' rate after rotational depletion of about 2 s$^{-1}$,
		which is also in line with the computed rates from the present, realistic calculations with respect to the larger Langevin rate employed 
		by the experiments.
\end{enumerate}

In conclusion, the present modeling of the collision-driven rotational cooling kinetics of MgH$^+$ ions trapped in a Coulomb crystal, 
and further exposed to the interaction with the uploaded He buffer gas, indicates this process to be rather efficient and to be within a characteristic
time  below  1 s. The internal rotational state populations are shown to reach thermalization with the  translational temperatures of the
buffer gas under the different trap conditions which are induced by the experimental tuning of the micromotion amplitudes. This is in near quantitative agreement
 with the experimental findings of ref. \cite{18} and agrees with
all the cooling dynamics features discussed by the experiments.

\section{Present conclusions}
\label{sec6}

The study reported in  this paper deals with the detailed computational modeling of the internal rotational state-changing 
kinetics of a molecular ion, the MgH$^+$($^1\Sigma$) ion, which , experimentally, is first undergoing sympathetic cooling in a Coulomb crystal trap
arrangement, and then is further internally cooled by collisions with an uploaded buffer gas of He atoms. By the tuning of its micromotion amplitudes that simulate the changing of its average relative collisional energy within the trap. In order to carry out a complete 
computational simulation from first principles, and using a quantum ab initio description of the various steps involved, we have 
obtained first the electronic potential energy surface for the interaction between MgH$^+$ and He atoms. To this aim, we have employed the set of ab initio points already
reported in our earlier work \cite{19,20} and have implemented them by generating additional points for the short-range regions of the 
repulsive part of the PES, as discussed and described in Section \ref{sec2}. The ensuing interaction potential has been used to calculate the partial,
integral, state-to-state inelastic cross sections between the lower five rotational states of the molecular ion, although only four
of them have been found to be significantly populated in the experiments. From the set of inelastic cross sections, which involve excitation
and de-excitation transitions between the rotational states, we have obtained the corresponding inelastic rates for the rotational  ''cooling'' and
the rotational ''heating'' collision-driven dynamics over a range of temperatures up to about 50 K, which is well above the experimentally tested trap temperatures in ref. \cite{18}
and also well above our earlier calculations on this same system \cite{19,20}. 

The solution of the Master Eq.s for the time evolution of the level populations during the uploading of the buffer gas allowed us to obtain 
quantitative estimates of the time interval needed to deplete the rotational states of the ion in order to thermalize its rotational state populations to the translational 
temperature of the trap after gas uploading. Our present results are in close  agreement with the experimental findings and suggest that:
\begin{enumerate}
	\item After about 1 s the internal energy distributions of the trapped ion have reached the translational temperature perceived after  the uploading of the buffer
		gas and achieved by the tuning of the  micromotion amplitude  around a fixed He temperature of  about 8.7 K. Our estimated ''refilling'' rate is of the order of about 
1-2 s$^{-1}$, a value which is close to the
		experimental estimates of about 1 s$^{-1}$ \cite{18}.
	\item The experimental estimate of a global cooling rate in the traps is around 10$\cdot$10$^{-10}$ cm$^3$s$^{-1}$, in line with 
		our dominant cooling rates at  around 20 K of about 6.6$\cdot$10$^{-10}$ cm$^3$s$^{-1}$.
	\item All the experimentally observed distributions between rotational states of the ion at different temperatures, are in near
		quantitative agreement with our thermalized rotational distributions after time intervals between 1 s and 3 s.
         \item The experimentally observed equalization between rotational and thermal temperatures of the ions in the trap are confirmed by our calculations, 
                 which report rotational  distributions at the various temperatures to be very close to the thermal distributions achieved after rapidly reaching the steady-state 
                  conditions in the traps.       
\end{enumerate}

The calculations have therefore found very good agreement with the experimental data and suggest that the collision-driven state-changing rates
for the present cations are indeed very large and indicate a very rapid process of thermalization of the rotational levels' ''temperature'' to 
the translational temperature  achieved   into the Coulomb crystal environment after the uploading of the He buffer gas and subsequent tuning of the ion micromotion.

\section{Acknowledgments}

L.G.S. acknowledges the financial support from the Spanish Ministry of Science and Innovation Grant No.  CTQ2015-65033-P.
F.A.G. and R.W. thank the support by the Austrian Science Fund (FWF), Project No. 29558-N36. 
The computational results have been obtained by using in-house computer codes running on the HPC infrastructure LEO of the University of Innsbruck.
This work was supported by a STSM Grant from COST Action CM1401, being held by L. Gonz{\' a}lez-S{\' a}nchez.
We thank I. Iskandarov and Lorenzo Petralia for their generous initial help in setting up the multipolar coefficients
from the computed interaction potential discussed in Section \ref{sec2}.

\section{Keywords}

Keyword 1: inelastic collisions, keyword 2: intermolecular potentials, keyword 3: collisionally inelastic rates, 
keyword 4: rotational relaxation times , keyword 5: molecular dynamics in cold traps



\bibliographystyle{unsrt}
\bibliography{mghphe.bib}

\begin{thebibliography}{10}

\bibitem{1}
William~D. Phillips.
\newblock Nobel lecture: Laser cooling and trapping of neutral atoms.
\newblock {\em Rev. Mod. Phys.}, 70:721--741, Jul 1998.

\bibitem{2}
e.g. see:~R. Krems, B.~Friedrich, and W.C. Stwalley.
\newblock {\em Cold Molecules: Theory, Experiment, Applications}.
\newblock CRC Press, 2009.

\bibitem{3}
Stefan Willitsch.
\newblock Coulomb-crystallised molecular ions in traps: methods, applications,
  prospects.
\newblock {\em Int. Rev. Phys. Chem.}, 31(2):175--199, 2012.

\bibitem{4}
Lincoln~D Carr, David DeMille, Roman~V Krems, and Jun Ye.
\newblock Cold and ultracold molecules: science, technology and applications.
\newblock {\em New Journal of Physics}, 11(5):055049, 2009.

\bibitem{5}
S.~Schiller and V.~Korobov.
\newblock Tests of time independence of the electron and nuclear masses with
  ultracold molecules.
\newblock {\em Phys. Rev. A}, 71:032505, Mar 2005.

\bibitem{6}
D.~DeMille.
\newblock Quantum computation with trapped polar molecules.
\newblock {\em Phys. Rev. Lett.}, 88:067901, Jan 2002.

\bibitem{7}
M.~{Shapiro} and P.~{Brumer}.
\newblock {\em Cold Molecules: Theory, Experiment, Applications}.
\newblock New York, Wiley, 2003.

\bibitem{8}
R.~V. Krems.
\newblock Cold controlled chemistry.
\newblock {\em Phys. Chem. Chem. Phys.}, 10:4079--4092, 2008.

\bibitem{9}
Daniel Hauser, Seunghyun Lee, Fabio Carelli, Steffen Spieler, Olga
  Lakhmanskaya, Eric~S. Endres, Sunil~S. Kumar, Franco Gianturco, and Roland
  Wester.
\newblock {Rotational state--changing cold collisions of hydroxyl ions with
  helium}.
\newblock {\em Nature Physics}, 11:467 EP --, May 2015.

\bibitem{10}
Matthieu Viteau, Amodsen Chotia, Maria Allegrini, Nadia Bouloufa, Olivier
  Dulieu, Daniel Comparat, and Pierre Pillet.
\newblock Optical pumping and vibrational cooling of molecules.
\newblock {\em Science}, 321(5886):232--234, 2008.

\bibitem{11}
Jonathan~D. Weinstein, Robert deCarvalho, Thierry Guillet, Bretislav Friedrich,
  and John~M. Doyle.
\newblock Magnetic trapping of calcium monohydride molecules at millikelvin
  temperatures.
\newblock {\em Nature}, 395:148 EP --, Sep 1998.

\bibitem{12}
Hendrick~L. Bethlem, Giel Berden, Floris M.~H. Crompvoets, Rienk~T. Jongma,
  Andr{\'e} J.~A. van Roij, and Gerard Meijer.
\newblock Electrostatic trapping of ammonia molecules.
\newblock {\em Nature}, 406:491 EP --, Aug 2000.

\bibitem{13}
T.~Rieger, T.~Junglen, S.~A. Rangwala, P.~W.~H. Pinkse, and G.~Rempe.
\newblock Continuous loading of an electrostatic trap for polar molecules.
\newblock {\em Phys. Rev. Lett.}, 95:173002, Oct 2005.

\bibitem{14}
R.~Fulton, A.~I. Bishop, and P.~F. Barker.
\newblock Optical stark decelerator for molecules.
\newblock {\em Phys. Rev. Lett.}, 93:243004, Dec 2004.

\bibitem{15}
K.~M\o{}lhave and M.~Drewsen.
\newblock {Formation of translationally cold ${\mathrm{MgH}}^{+}$ and
  ${\mathrm{MgD}}^{+}$ molecules in an ion trap}.
\newblock {\em Phys. Rev. A}, 62:011401, Jun 2000.

\bibitem{16}
S.~Trippel, J.~Mikosch, R.~Berhane, R.~Otto, M.~Weidem\"uller, and R.~Wester.
\newblock {Photodetachment of Cold ${\mathrm{OH}}^{\ensuremath{-}}$ in a
  Multipole Ion Trap}.
\newblock {\em Phys. Rev. Lett.}, 97:193003, Nov 2006.

\bibitem{17}
Peter~F. Staanum, Klaus H{\o}jbjerre, Peter~S. Skyt, Anders~K. Hansen, and
  Michael Drewsen.
\newblock Rotational laser cooling of vibrationally and translationally cold
  molecular ions.
\newblock {\em Nature Physics}, 6:271, Mar 2010.

\bibitem{18}
A.~K. Hansen, O.~O. Versolato, L.~Klosowski, S.~B. Kristensen, A.~Gingell,
  M.~Schwarz, A.~Windberger, J.~Ullrich, J.~R.~Crespo L{\'o}pez-Urrutia, and
  M.~Drewsen.
\newblock Efficient rotational cooling of coulomb-crystallized molecular ions
  by a helium buffer gas.
\newblock {\em Nature}, 508:76, Mar 2014.

\bibitem{19}
M.~Tacconi, F.~A. Gianturco, E.~Yurtsever, and D.~Caruso.
\newblock {Cooling and quenching of $^{24}\mathrm{Mg}$H${}^{+}$($X
  {}^{1}{\ensuremath{\Sigma}}^{+}$) by $^{4}\mathrm{He}$(${}^{1}S$) in a
  Coulomb trap: A quantum study of the dynamics}.
\newblock {\em Phys. Rev. A}, 84:013412, Jul 2011.

\bibitem{20}
Domenico Caruso, Mario Tacconi, Franco~A. Gianturco, and Ersin Yurtsever.
\newblock {Quenching vibrations by collisions in cold traps: A quantum study
  for MgH$^+$(X$^1\Sigma^+$) with 4He($^1$S)}.
\newblock {\em Journal of Chemical Sciences}, 124(1):93--97, Jan 2012.

\bibitem{21}
M.~J. Frisch, G.~W. Trucks, H.~B. Schlegel, G.~E. Scuseria, M.~A. Robb, J.~R.
  Cheeseman, J.~A. Montgomery, Jr., T.~Vreven, K.~N. Kudin, J.~C. Burant, J.~M.
  Millam, S.~S. Iyengar, J.~Tomasi, V.~Barone, B.~Mennucci, M.~Cossi,
  G.~Scalmani, N.~Rega, G.~A. Petersson, H.~Nakatsuji, M.~Hada, M.~Ehara,
  K.~Toyota, R.~Fukuda, J.~Hasegawa, M.~Ishida, T.~Nakajima, Y.~Honda,
  O.~Kitao, H.~Nakai, M.~Klene, X.~Li, J.~E. Knox, H.~P. Hratchian, J.~B.
  Cross, V.~Bakken, C.~Adamo, J.~Jaramillo, R.~Gomperts, R.~E. Stratmann,
  O.~Yazyev, A.~J. Austin, R.~Cammi, C.~Pomelli, J.~W. Ochterski, P.~Y. Ayala,
  K.~Morokuma, G.~A. Voth, P.~Salvador, J.~J. Dannenberg, V.~G. Zakrzewski,
  S.~Dapprich, A.~D. Daniels, M.~C. Strain, O.~Farkas, D.~K. Malick, A.~D.
  Rabuck, K.~Raghavachari, J.~B. Foresman, J.~V. Ortiz, Q.~Cui, A.~G. Baboul,
  S.~Clifford, J.~Cioslowski, B.~B. Stefanov, G.~Liu, A.~Liashenko, P.~Piskorz,
  I.~Komaromi, R.~L. Martin, D.~J. Fox, T.~Keith, M.~A. Al-Laham, C.~Y. Peng,
  A.~Nanayakkara, M.~Challacombe, P.~M.~W. Gill, B.~Johnson, W.~Chen, M.~W.
  Wong, C.~Gonzalez, and J.~A. Pople.
\newblock Gaussian 03, \uppercase{R}evision \uppercase{C}.02.
\newblock \uppercase{G}aussian, Inc., Wallingford, CT, 2004.

\bibitem{22}
A.~Stone.
\newblock {\em The Theory of Intermolecular Forces}.
\newblock The Theory of Intermolecular Forces. OUP Oxford, 2013.

\bibitem{23}
e.g. see:~J.R. Taylor.
\newblock {\em Scattering Theory: The Quantum Theory of Nonrelativistic
  Collisions}.
\newblock Dover Books on Engineering. Dover Publications, 2012.

\bibitem{24}
R.~Martinazzo, E.~Bodo, and F.A. Gianturco.
\newblock A modified variable-phase algorithm for multichannel scattering with
  long-range potentials.
\newblock {\em Computer Physics Communications}, 151(2):187 -- 198, 2003.

\bibitem{25}
D.~L{\'o}pez-Dur{\'a}n, Enrico Bodo, and Franco~A Gianturco.
\newblock {ASPIN}: An all spin scattering code for atom-molecule
  rovibrationally inelastic cross sections.
\newblock {\em Computer Physics Communications}, 179:821--838, 2008.

\bibitem{26}
L.~Gonz\'alez-S\'anchez, F.~A. Gianturco, and R.~Wester.
\newblock {State--changing processes for ions in cold traps: LiH$^-$ molecules
  colliding with He as a buffer gas}.
\newblock {\em Journal of Physics B: Atomic, Molecular and Optical Physics},
  49(23):235201, 2016.

\bibitem{27}
Mario~Hern\'andez Vera, F.~A. Gianturco, R.~Wester, H.~da~SilvaJr., O.~Dulieu,
  and S.~Schiller.
\newblock {Rotationally inelastic collisions of H$_2^+$ ions with He buffer
  gas: Computing cross sections and rates}.
\newblock {\em The Journal of Chemical Physics}, 146(12):124310, 2017.

\bibitem{30}
Jes\'us P\'erez-R\'{\i}os and F.~Robicheaux.
\newblock Rotational relaxation of molecular ions in a buffer gas.
\newblock {\em Phys. Rev. A}, 94:032709, Sep 2016.

\bibitem{31}
L.~Gonz\'alez-S\'anchez, F.A. Gianturco, and R.~Wester.
\newblock {Modeling quantum kinetics in ion traps: state--changing collisions
  for OH$^+$($^3\Sigma^+$) ions with He as a buffer gas}.
\newblock {\em ChemPhysChem}, 19:1866, 2018.

\bibitem{28}
S.~Schiller, I.~Kortunov, M.~Hern\'andez~Vera, F.~Gianturco, and H.~da~Silva.
\newblock {Quantum state preparation of homonuclear molecular ions enabled via
  a cold buffer gas: An ab initio study for the ${\mathrm{H}}_{2}{}^{+}$ and
  the ${\mathrm{D}}_{2}{}^{+}$ case}.
\newblock {\em Phys. Rev. A}, 95:043411, Apr 2017.

\bibitem{29}
Mario~Hern{\'a}ndez Vera, Stephan Schiller, Roland Wester, and
  Francesco~Antonio Gianturco.
\newblock {Rotationally inelastic cross sections, rates and cooling times for
  para-H$_2^+$, ortho-D$_2^+$ and HD$^+$ in cold helium gas}.
\newblock {\em Eur. Phys. J. D}, 71(5):106, May 2017.

\end{thebibliography}

\end{document}